\documentclass[twocolumn,showpacs,preprintnumbers,amsmath,amssymb]{revtex4-2}

\usepackage{CJK}

\usepackage{graphicx}

\usepackage{dcolumn}

\usepackage{bm}
\usepackage[normalem]{ulem}
\usepackage{color}

\usepackage{epstopdf}

\def\rf#1{(\ref{#1})}
\newcommand{\beps}{\bm{\epsilon}}
\newcommand{\bew}{\begin{widetext}}
\newcommand{\ew}{\end{widetext}}

\newcommand{\dw}{\delta\omega}
\newcommand{\nn}{\nonumber}

\newcommand{\ii}{{\rm i}}

\newcommand{\bx}{\mathbf{x}}
\newcommand{\by}{\mathbf{y}}

\newcommand{\bv}{\mathbf{v}}

\newcommand{\br}{\mathbf{r}}
\newcommand{\bR}{\mathbf{R}}
\newcommand{\bff}{\mathbf{f}}

\newcommand{\hx}{\bf{\hat{x}}}

\newcommand{\sep}{ \ \ \ , \ \ \ }

\newcommand{\beq}{\begin{equation}}
\newcommand{\eeq}{\end{equation}}
\newcommand{\beqn}{\begin{eqnarray}}
\newcommand{\eeqn}{\end{eqnarray}}
\newcommand{\pp}{\partial}
\newcommand{\dd}{{\rm d}}
\newcommand{\ee}{{\rm e}}

\newcommand{\la}{\langle}

\newcommand{\ra}{\rangle}

\begin{document}

\begin{CJK*}{GBK}{}
%\preprint{APS/123-QED}

\title{Only the Ambidextrous Can Flock: Two-dimensional Chiral Malthusian Flocks,  Time Cholesterics, and the KPZ Equation}
\author{Leiming Chen}
	\email{leiming@cumt.edu.cn}
	\affiliation{School of Materials science and Physics, China University of Mining and Technology, Xuzhou Jiangsu, 221116, P. R. China}
	\author{Chiu Fan Lee}
	\email{c.lee@imperial.ac.uk}
	\affiliation{Department of Bioengineering, Imperial College London, South Kensington Campus, London SW7 2AZ, U.K.}
	\author{John Toner}
	\email{jjt@uoregon.edu}
	\affiliation{Department of Physics and  Institute for Fundamental
 Science, University of Oregon, Eugene, OR $97403$}
	
\date{\today}
\begin{abstract}
We study two-dimensional chiral dry Malthusian flocks; that is, chiral polar-ordered active matter with neither number nor momentum conservation.
In the absence of fluctuations, these form a ``time cholesteric", in which the velocity rotates uniformly in {\it time} at a fixed frequency.
Fluctuations are described by the
($2+1$)-Kardar-Parisi-Zhang (KPZ) equation, which implies short-ranged orientational order.  For weak chirality,  the  system is in  
 the linear regime of the KPZ equation for a wide range of length scales, over which it exhibits quasi-long-ranged orientational order. 
 Our predictions for velocity and density correlations are testable in both simulations and experiments.
\end{abstract}
%\pacs{05.65.+b, 64.60.Ht, 87.18Gh}
\maketitle
\end{CJK*}

%\section{Introduction}{\label{Intro}}

Active matter \cite{book, Active1, Active2, Vicsek, TT1, TT3, birdrev, Active3, Active4, tissue, PS, Nematics, Chate1, Chate2} differs from equilibrium systems due to a variety of intrinsically non-equilibrium effects, including self-propulsion \cite{book, Vicsek, TT1, TT3, birdrev} and birth and death \cite{Toner_prl12, Chen_prl20, Chen_pre20}. These non-equilibrium effects, which are most commonly found in biological systems, lead to many surprising phenomena, including potential long-ranged order in two-dimensional (2D) systems with spontaneously broken continuous symmetries \cite{book, Vicsek, TT1, TT3, birdrev, inconvenienttruth} (i.e., 2D flocks), and hydrodynamic instabilities in very viscous systems \cite{sppprl, activerheo}.

While it is not an intrinsically non-equilibrium phenomenon, chirality \cite{glove} is also a ubiquitous feature of biological systems. It has recently been shown \cite{Liebchen_prl2017} that chirality in a collection of self-propelled particles (hereafter ``flockers") moving in two dimensions - i.e., a 2D ``flock"- can lead to a very unusual ``steady-state", in which the flockers move coherently in circles. That is, at any instant of time, all of the flockers are moving coherently in the same direction, but that coherent direction rotates at a constant angular velocity.
As a result, each flocker moves in a circle around a center unique to itself, in synchrony with all of the other flockers.

\begin{figure}
		\begin{center}
	\includegraphics[scale=.3]{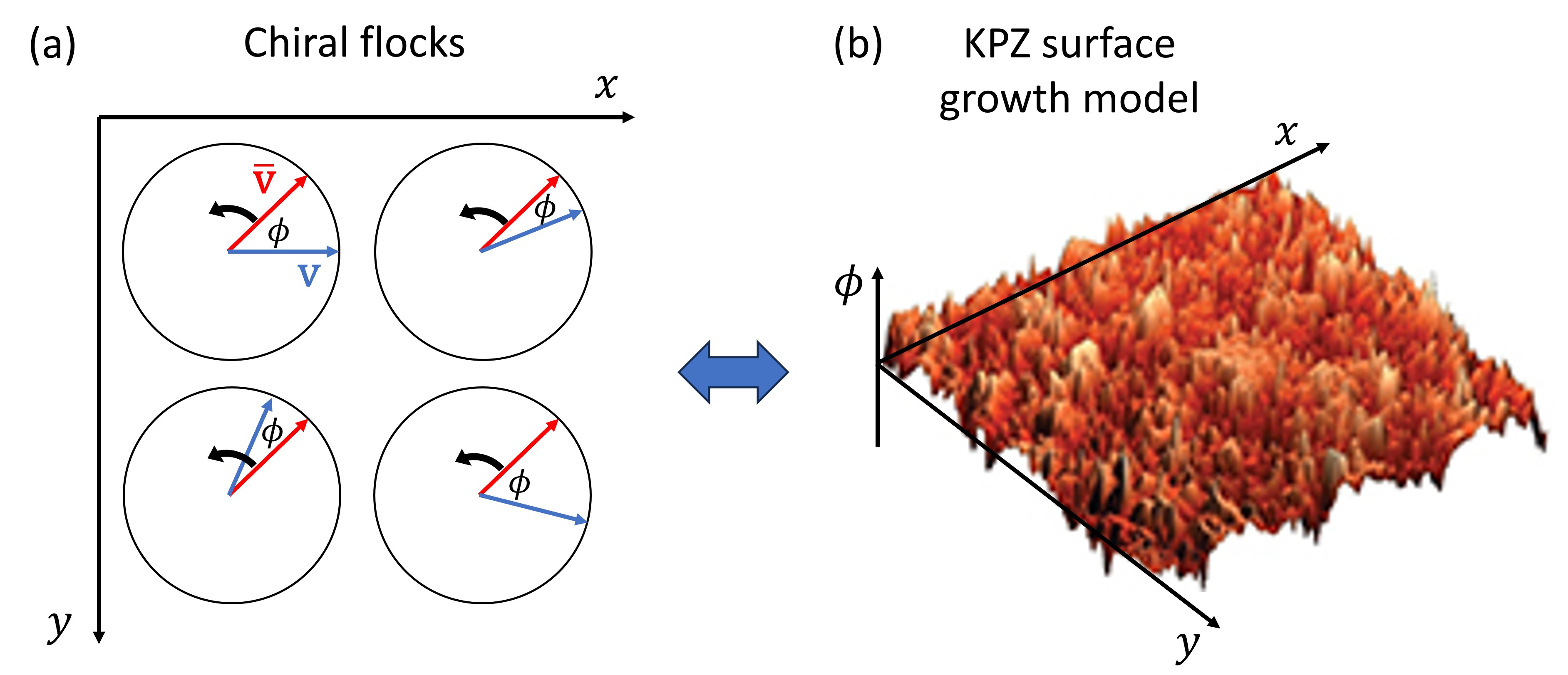}
		\end{center}
		\caption{(a) In a generic 2D chiral Malthusian flock, the mean velocity of the whole flock is denoted by $\bar{\bv}$ (red) , where $\bar{\bv}=v_0[\cos(bt)\hat{\bf x}-\sin(bt)\hat{\bf y}]$, while the spatio-temporally fluctuating velocity field is denoted by $\bv$ (blue) where $\bv=v_0[\cos(bt+\phi)\hat{\bf x}-\sin(bt+\phi)\hat{\bf y}]$. Ignoring variations in $|\bv|$, all fluctuations can be captured by the angular field $\phi(x,y,t)$. (b) Here, we show that the hydrodynamic behavior of $\phi$ can generically be mapped on the Kardar-Parisi-Zhang surface growth model \cite{KPZ} in (2+1) dimensions. 
The image in (b) is reproduced with permission from EPL {\bf 109}, 46003 (2015) \cite{cartoon}.
%\sout{		(b) [figure adapted from \cite{cartoon}].}
		}
		\label{cartoon}
	\end{figure}

Mathematically, the coarse-grained velocity field $\bv(\br,t)$ of a chiral flock in this state is spatially uniform (that is, independent of 2D position $\br$), and periodic in time:
\beq
\bv=v_0[\cos(bt+\phi)\hat{\bf x}-\sin(bt+\phi)\hat{\bf y}] \,,%,,~~~ \rho=\rho_0,
\label{SSI}
\eeq
where $b$ is an intrinsic frequency characteristic of the system. An achiral system must have $b=0$, since, without chirality, it has no way to decide whether to rotate right or left.
 Hence, we can use $b$ as a measure of the chirality of the system.

% However, it should be noted that, in general, there is no unique measure of the chirality of a chiral system. Hence, it is therefore possible that, by tuning one parameter of a chiral flock (e.g., by mixing ``left-handed" particles into a ``right-handed" flock), one might tune the rotation rate $b$ to zero {\it without} making other chiral parameters vanish. A similar phenomenon  happens \cite{glove} in equlibrium chiral nematic liquid crystals (i.e., ``cholestrics"), in which the coefficient of the ``twist" term  $\hn\cdot(\nabla\times\hn)$ in the Frank free energy of such systems can be made to vanish, by, e.g., mixing together two species of molecules of  opposite handedness that are not mirror images of each other, while the coefficients of  higher order chiral terms like, e.g.,  $[\hn\cdot(\nabla\times\hn)]^3$, do not vanish. 
 
%In the cholesteric case, however, {\it all} chiral terms can be made to vanish if one makes a ``racemic" mixture, in which two species of chiral molecules that are mirror images of each other, are mixed together. When such a mixture reaches $50-50$, one has obviously restored chiral symmetry to the system, and so, {\it all} chiral coefficients must vanish at precisely the same $50-50$ concentration.
 
 In our discussion of low chirality systems in this paper, we will assume that the chirality is lowered by  ``racemic mixing'' of chiral components that are perfect mirror images of each other. In such systems, all chiral coefficients vanish at the same concentration. Other possibilities will be discussed in \cite{ALP}. 

Note also that any ``snapshot" of such a flock (i.e., a flock with its velocity field given by \rf{SSI}) has perfect infinite ranged orientational order; that is, at every instant of time, all of the particles are moving in precisely the same direction.

Of course, in a time-averaged sense, rotation invariance is {\it not} broken in a state like \rf{SSI}; all directions of velocity are sampled for equal time.  The symmetry that {\it is} broken by \rf{SSI} is {\it time-translation invariance}{, which is broken because the system spontaneously chooses a particular phase $\phi$ in \rf{SSI}, which is equivalent to choosing a particular set of discrete times for the velocity to point, say, along $\hx$. In this sense, the state \rf{SSI} can be thought of as a type of ``time cholesteric'': the velocity rotates in {\it time} in the same way that the molecular axis rotates in {\it space} in a cholesteric liquid crystal. That is, the active chiral flock can thought of as a cholesteric with the ``pitch axis" being {\it time}\cite{Ananyo}.

One very simple and natural way to simulate a chiral flock is to modify the ``Vicsek" algorithm by altering the direction selecting step of the algorithm to make the particles select, not the average direction of its neighbors, but a direction that is, at all times and for all flockers, some fixed angle $\delta$ to the right of that mean direction (obviously, by choosing $\delta$ negative, one can make the flockers turn consistently to the left). Indeed, such simulations have been done \cite{Chate_chiral} , as have simulations of somewhat different chiral models \cite{Liebchen_epl2022, Liebchen_prl2017}, and do, indeed, find a state like \rf{SSI}.

% Note, however, that in a time averaged sense, this state of the system does not break rotation invariance (all directions of v equally likely if you wait a long time (t>>1/b). Symmetry that broken is actually time translation invariance. LRO=``synchronization``. (ref Grinstein)

In this paper, and the Associated Long Paper (ALP)\cite{ALP}, we
 investigate whether this long-ranged order is stable in the presence of noise, and furthermore determine the nature, size, and scaling of the fluctuations induced by the noise. In particular, we focus on 2D chiral ``dry Malthusian'' flocks - that is, flocks in which neither momentum or flocker number are conserved. The absence of conservation of momentum is due to the presence of a frictional substrate over which the self-propelled flockers move, while the absence of number conservation is a consequence of birth and death of flockers while in motion.

% Note that the solutions \rf{SSI} differ trivially by a constant phase $\phi$.

{\it Malthusian chiral active fluids = KPZ  equation---}We first formulate the generic equation of motion (EOM) for chiral Malthusian flocks by allowing for chirality into the standard Malthusian EOM \cite{Toner_prl12}, which in two dimensions amounts to incorporating the rank-2 anti-symmetric tensor $\beps$: 
\beq
\bm{\epsilon} =
\left(\begin{array}{rr}
	0 & 1
	\\
	-1 & 0
	\end{array}
\right)
\eeq
into the achiral EOM {\it wherever possible}. 
Here, we focus on a ``truncated" version of our generic EOM as its study already illustrates   the fact that this problem maps onto the $2+1$-dimensional KPZ equation. In \cite{ALP} we show that this conclusion also holds for the most general possible model. Our truncated EOM is
\bew
\beqn
\pp_t \bv +\lambda_1 (\bv \cdot {\bf \nabla})\bv +\lambda'_1(\bv \cdot \bm{\epsilon} \cdot {\bf \nabla})\bv+ \lambda_2(\nabla\cdot\bv)\bv+\lambda_3\nabla(v^2)
%+\lambda_3 \bm{\epsilon} \cdot \left[(\bv \cdot {\bf \nabla})\bv\right]
&=&\mu_1 \nabla^2 \bv+\mu_2\nabla(\nabla\cdot\bv)+\mu_3(\bv\cdot\nabla)^2\bv+\mu' \bm{\epsilon}\cdot \nabla^2 \bv\nonumber\\
&&+\left[L(|\bv|)+b\bm{\epsilon}\cdot\right]\bv +\bff
\ ,\label{EOM:chiral_1}
\eeqn
\ew
where $\bv(\br,t)$ is the local velocity field, $L(|\bv|)$ is a Lagrange multiplier that enforces a ``fixed length`` or ``constant speed'' constraint on the velocity that $|\bv|=v_0={\rm constant}\ne0$ everywhere. In this chiral system, $b>0$ if the particles tend to turn right, while $b<0$
if the particles tend to turn left. Further,   the random force $\bff$ here can also be chiral in principle  but  its  spatio-temporal {\it statistics} are {\it achiral}:  
\beq
\la f_m (\br,t) f_n(\br',t') \ra =2D \delta_{mn}\delta^2(\br-\br')\delta(t-t')
\ .
\eeq

Here, we focus on the flocking phase, whose noiseless state is described by Eq.~(\ref{SSI}). 
To study the effects of noise, we allow  the phase  $\phi$ in \rf{SSI} to vary  in space and time; i.e.,  we take the velocity to be
\beq
\bv(\br,t)=v_0\left[\cos(bt+\phi(\br,t))\hat{\bx}-\sin(bt+\phi(\br,t))\hat{\by}\right]\,.
%\,,~~~ \rho=\rho_0+\delta\rho(\br,t)\,.
 \label{v_fluctuate}
\eeq
Inserting (\ref{v_fluctuate}) into EOM (\ref{EOM:chiral_1}),
%\sout{and adding a Lagrangian multiplier $L(|\bv|)\bv$ to (\ref{EOM:chiral_1}) to enforce the constant-magnitude restriction on $\bv$,}
we get, in tensor form,
\beq
\pp_tv_i=(b+\pp_t\phi)\epsilon_{ik}v_k \sep \pp_jv_i=\epsilon_{ik}v_k\pp_j\phi
\,.
\label{vderivs}
\eeq

Using these in \rf{EOM:chiral_1}, we obtain the EOM for the phase $\phi$:
\bew
\beqn
\epsilon_{ij}v_j\pp_t \phi
%+\lambda_3 \bm{\epsilon} \cdot \left[(\bv \cdot {\bf \nabla})\bv\right]
&=&-\lambda_1 \epsilon_{ik}v_kv_j\pp_j\phi-\lambda'_1\left[v_i(v_k\pp_k\phi)-v_0^2(\pp_i\phi)\right] -\lambda_2\epsilon_{jk}v_iv_k(\pp_j\phi) +\mu_1\left[\epsilon_{ik}v_k\pp_j\pp_j\phi-v_i(\pp_j\phi)(\pp_j\phi)\right]\nonumber\\
&&+\mu_2\left[\epsilon_{jk}v_k\pp_i\pp_j\phi-v_j(\pp_i\phi)(\pp_j\phi)\right]
+\mu_3\left[\epsilon_{i\ell}v_\ell v_jv_k\pp_j\pp_k\phi-v_iv_kv_j(\pp_k\phi)(\pp_j\phi)\right]
\nonumber\\&&-\mu'\left[v_i\pp_k\pp_k\phi+\epsilon_{im}v_m(\pp_k\phi)(\pp_k\phi)\right]+L(|v|)v_i +f_i
\ .\label{EOM:chiral_3}
\eeqn
\ew
In the low-noise limit (i.e., deep into the flocking phase), we expect $\phi$ to vary slowly in time. 
To calculate the slow evolution of $\phi$, we first multiply both sides of (\ref{EOM:chiral_3}) by $\epsilon_{in}v_n$,  with the usual implied summation convention on the repeated index $n$. Doing so eliminates all terms along $\bv$ in (\ref{EOM:chiral_3}) (that is, in tensor notation, all terms proportional to $v_i$, which includes the Lagrange multiplier term $L(|v|)v_i $, along with several others). We are thus left with
\beqn
v_0^2\pp_t \phi
&=&-\lambda_1 v_0^2v_j\pp_j\phi+\lambda'_1v_0^2\epsilon_{in}v_n\pp_i\phi +\mu_1v_0^2\nabla^2\phi
\nonumber\\
&&
+\mu_2\left[v_0^2\nabla^2\phi-v_iv_ j\pp_i\pp_j\phi-\epsilon_{in}v_nv_j(\pp_i\phi)(\pp_j\phi)\right]
\nonumber\\
&&+\mu_3v_0^2 v_jv_k\pp_j\pp_k\phi-\mu'v_0^2|\nabla\phi|^2 +\epsilon_{in}v_nf_i
\ ,\label{EOM:chiral_4}
\eeqn
where we have made liberal use of the following identities obeyed by the rank-2 anti-symmetric tensor $\beps$:
\beq
\epsilon_{ij}\epsilon_{kl}=\delta_{ik}\delta_{jl}-\delta_{il}\delta_{jk}\,,~~~ \epsilon_{ik}\epsilon_{jk}=\delta_{ij} \,.
\label{epsid}
\eeq
Second, we time-average Eq.~(\ref{EOM:chiral_4})  over  one  oscillation cycle of $\bv$. For the purposes of such time averages,  $\phi(\br,t)$ and all of its spatio-temporal derivatives can be treated as constants, since $\phi(\br,t)$ varies slowly on the time scale $b^{-1}$ of the oscillating velocity.
They can therefore be removed from the averages over one cycle, which we'll denote by $\langle\rangle_c$. It is then fairly straightforward to show that time averages of products of two and four components of $\bv$ are given by:
\beqn
\begin{aligned}
&\langle v_iv_j\rangle_c={v_0^2\delta_{ij}\over2} \ \ , \\
&\langle v_iv_jv_kv_l\rangle_c={v_0^4\over8}\bigg(\delta_{ij}\delta_{kl}+\delta_{ik}\delta_{jl}+\delta_{il}\delta_{jk}\bigg) \,.
\end{aligned}
\label{angaves}
\eeqn
Averages over a full cycle of products of any odd number of components of $\bv$ clearly vanish. This immediately eliminates all of the $\lambda$-terms, since they have an odd number of components of $\bv$. This is one of the radical changes introduced by the chirality, since those terms are known   to dominate the physics of the achiral problem \cite{Toner_prl12, Chen_prl20,Chen_pre20}.

Using \rf{angaves} on the average of (\ref{EOM:chiral_4}) finally
gives the KPZ equation as promised in the abstract:
\beqn
\pp_t\phi=\nu\nabla^2\phi+{\lambda_{_K}\over 2}(\nabla\phi)^2+f_\phi\,,\label{EOM:phi}
\eeqn
where $\nu=\mu_1+\left({\mu_2+\mu_3v_0^2\over 2}\right)$, $\lambda_{_K}=-\mu^\prime$ and
 where  $f_\phi$ is a Gaussian, zero-mean white noise with the following statistics:
\beq
\la f_\phi (\br,t) f_\phi(\br',t') \ra =2D_\phi \delta^2(\br-\br')\delta(t-t')
\ ,
\label{fphicorr}
\eeq
with $D_\phi\equiv(D/v_0^2)$.
A schematic representation of the mapping is shown in Fig.~\ref{cartoon}. 

We note that the KPZ universality class (UC) encompasses an amazing array of diverse systems--recent additions to this include incompressible polar active fluids \cite{chen_natcomm16}, driven Bose-Einstein condensates \cite{diessel_prl22}, and non-reciprocal systems \cite{pisegna_pnas24,daviet_a24}.
%[CFL: Here's where I added some more recent papers on KPZ]. 
It is nonetheless  surprising, in our opinion, that all generic 2D chiral Malthusian flocks   necessarily fall into the KPZ UC as well.

}

{\it Consequences of the mapping to the KPZ equation---}
As is well known \cite{KPZ}, the fluctuations of a field $\phi$ described by the (2+1)-KPZ equation diverge in the limit of large distances. 
Specifically, 
\begin{eqnarray}
\label{phi_correlI_phi}
\langle[\phi(\br,t)-\phi(\br',t^\prime)]^2\rangle=\left\{
\begin{array}{ll}
A_\phi|\br-\br'|^{2\chi}\,,&{B_\phi|t-t^\prime|\over|\br-\br'|^z}\ll 1\,,\\
%|\br-\br'|
A_\phi|t-t^\prime|^{2\chi\over z}\,,&{B_\phi|t-t^\prime|\over|\br-\br'|^z}\gg1\,,
\end{array}
\right.\nn\\
\end{eqnarray}
where
%\sout{$\chi$ is the roughness exponent for (2+1)-KPZ equation}
$A_\phi$ and $B_\phi$ are non-universal positive constants.

The current estimate of the values of
%\sout{most accurate numerical result for}
$\chi$ and $z$ based on simulations are \cite{kpzexp1, kpzexp2,kpzexp3, kpzexp4,kpzexp5, kpzexp6} $\chi=0.388\pm.002$,  $z=1.622\pm.002$.

Since $\chi>0$, the equal-time correlation in $\phi$ diverges with distance, which 
%This divergence \rf{phi_correlI_v0} 
implies that  equal-time velocity correlations will be short-ranged. That is, chirality destroys the long-ranged orientational order present \cite{Toner_prl12,Chen_prl20,Chen_pre20} in an achiral  Malthusian flock. This is our principal conclusion.

In fact, chiral Malthusian flocks are probably even more disordered  than this would suggest. This is because, in contrast to the usual KPZ equation, the achiral Malthusian flock  maps onto the {\it compact} $(2+1)$-dimensional KPZ equation.

What we mean by ``compact''  is that, unlike, say, a growing interface, in which the KPZ variable $\phi$ is the height of the interface, and every value of the height represents a physically distinct local state, our KPZ variable is a {\it phase}. That is, a state with a given local value $\phi(\br,t)$ is locally physically indistinguishable from state $\phi(\br,t)+2\pi n$, for any integer $n$, as can be seen directly from Eq.~\rf{v_fluctuate}.
%\rf{v_fluctuateI}.

This is the same symmetry that is present in the 2D XY model. Here, as there, it allows   topological defects -i.e., vortices \cite{KT}.  These become important on length scales larger than the mean inter-vortex distance  $L_v$, and  destroy the order even more thoroughly than the ``spin-wave" (i.e., vortex free) fluctuations embodied in \rf{phi_correlI_phi}. 

The exact behavior of vortices in the compact KPZ equation remains an open question \cite{Ehud}. 
In another active system - namely,  active smectics \cite{apolar Malthusian}, topological defects behave very differently from those in their equilibrium counterparts \cite{JPJTFJ}. Our results here for the vortex length $L_v$ and the behavior of a compact KPZ equation on length scales longer than $L_v$ are based upon our speculation that this does {\it not} happen in KPZ like systems. We believe this is reasonable, for reasons we'll discuss in the ALP \cite{ALP}; however, our conclusions about vortices   are somewhat speculative.

{\it Small chirality limit---}When the chirality $b$ is small, the system looks, over a large range of length and time scales, as if it were {\it achiral}. Over these length and time scales, a weakly chiral system will exhibit ``anomalous hydrodynamics'', with its effective damping coefficients growing with length scale. This leads, as we show in detail in the  (ALP) \cite{ALP},  to a strong dependence of
the parameters $\nu$ and $\lambda_{_K}$ in our KPZ equation on the chirality $b$; they  respectively diverge and vanish in the limit of small  $b$;  for the ``racemic'' case described above
\beq
\nu(b)\propto b^{-\eta_\nu} \sep \lambda_{_K}(b)\propto b \,,
\label{mubI}
\eeq
where the universal exponent $\eta_\nu$ is related to the dynamical exponent \cite{Toner_prl12,Chen_prl20} $z_a$ of ``achiral`` Malthusian flocks via the exact relation
\beq
\eta_\nu={2\over z_a}-1\approx{3\over 5} \,.
\label{etanu}
\eeq
The numerical value given here is based on numerical  simulations of the noisy hydrodynamic equation for achiral Malthusian flocks by  Ref.~\cite{Chate_prl24}. 

The other parameter in the KPZ equation, which is the noise strength $D_\phi$, is independent of chirality $b$ in the limit of small chirality.

\begin{figure}
		\begin{center}
			\includegraphics[scale=.26]{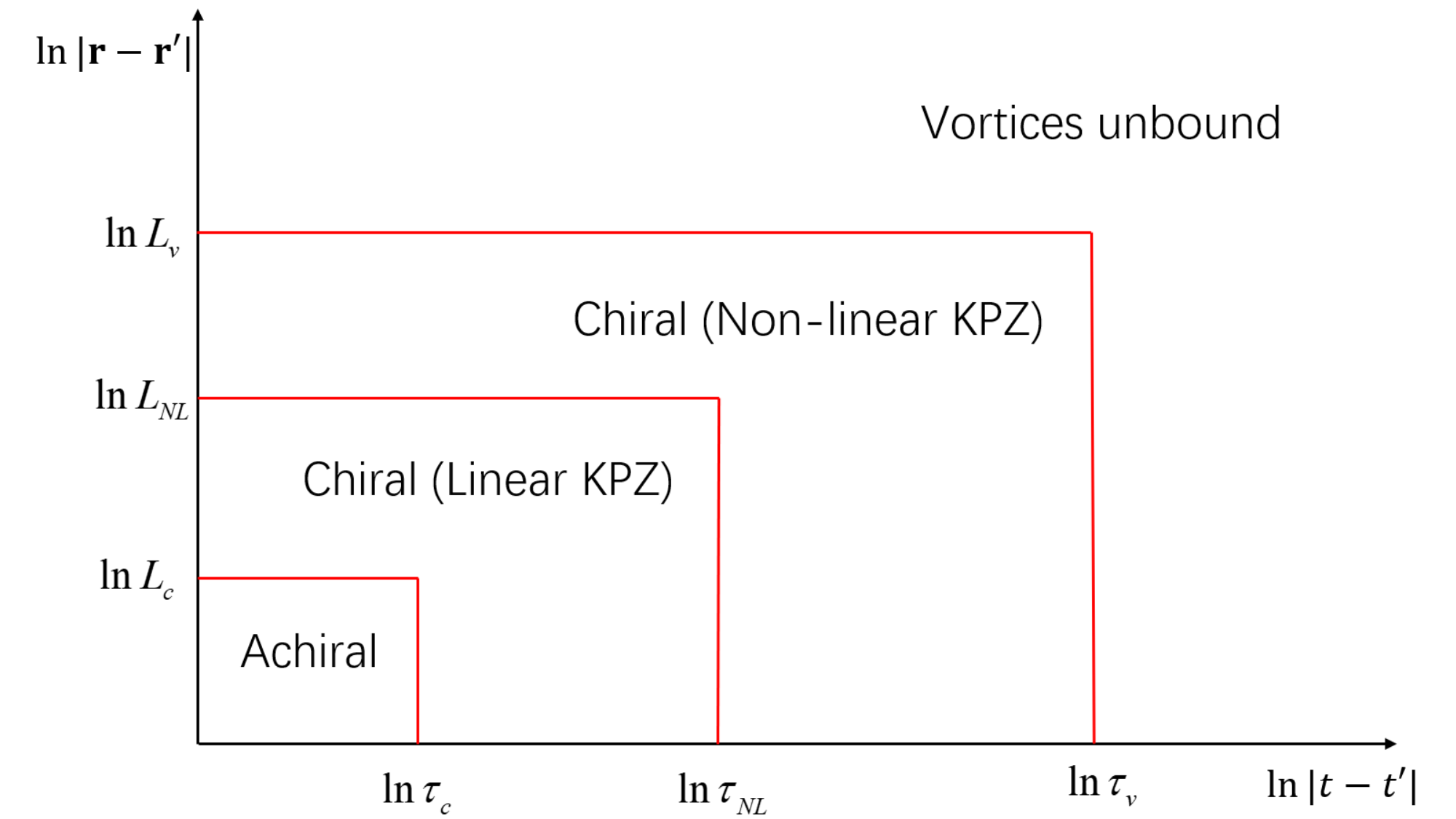}
		\end{center}
		\caption{ Regimes of different behavior in the limit of weak chirality. The longest length scales $L_{_{NL}}$ and $L_v$ and time scales $\tau_{_{NL}}$ and $\tau_v$ will be long enough to allow several decades of length and time scale to lie in each region, making our scaling predictions for each region experimentally accessible. See the text for more details.}
		\label{ltsc}
	\end{figure}

Despite our fundamental conclusion that chirality destroys long-ranged order in 2D Malthusian chiral flocks, for {\it small} chirality $b$, the aforementioned strong divergence of $\nu(b)$   and vanishing of $\lambda$ as $b\to0$ implies that order persists out to quite large distances. In fact, we find, as shown in detail in the  (ALP) \cite{ALP},  that for weak chirality, there is a hierarchy of length and time scales, illustrated in
Fig.~\ref{ltsc}, separating regimes of quite different  scaling behavior of the velocity correlations $\langle\bv(\br, t)\cdot\bv(\br^\prime, t^\prime)\rangle$. Some of these regions will be effectively ordered.

All of these length and time scales diverge in the limit of small chirality $b$, according to the following scaling laws:
\begin{align}
L_c(b) &\propto b^{-1/z_a} \sep 1/z_a\approx4/5 \,,
\label{lchirI}
\\
L_{_{NL}}(b)&\propto \exp\left(C_{_{NL}}b^{-\eta_g}\right)\ , \ \eta_g=6/z_a-1\approx19/5 \,,
\label{lnlI}
\\
L_v(b)&\propto \exp\left(C_{vL}b^{-\eta_y}\right)\ , \ \eta_y=8/z_a-2\approx22/5 \,, 
\label{lvI}
\\
\tau_c(b)&\propto  b^{-1} \,,
\label{taucI}
\\
\tau_{_{NL}}(b)&\propto \exp\left(2C_{_{NL}}b^{-\eta_g}\right) \,,
\label{taunlI}
\\
\tau_v(b) &\propto  \exp\left( C_{v\tau}b^{-\eta_y}\right) \,,
\label{tauvI}
\end{align}
%
%
%
%\beq
%L_c(b)\propto b^{-1/z_a} \sep 1/z_a\approx4/5 \,,
%\label{lchirI}
%\eeq
%
%\beq
%L_{_{NL}}(b)\propto \exp\left(C_{_{NL}}b^{-\eta_g}\right)  \sep \eta_g\approx19/5 \,,
%\label{lnlI}
%\eeq
%
%
%\beq
%L_v(b)\propto \exp\left(C_{vL}b^{-\eta_y}\right)  \sep \eta_y\approx22/5 \,,
%\label{lvI}
%\eeq
%
%\beq
%\tau_c(b)\propto b^{-1} \,,
%\label{taucI}
%\eeq
%
%\beq
%\tau_{_{NL}}(b)\propto \exp\left(2C_{_{NL}}b^{-\eta_g}\right) \,,
%\label{taunlI}
%\eeq
%
%
%\beq
%\tau_v(b)\propto \exp\left( C_{v\tau}b^{-\eta_y}\right) \,,
%\label{tauvI}
%\eeq
where $C_{_{NL}}$, $C_{vL}$,  and $C_{v\tau}$ are all non-universal constants.

We will now describe the behavior of the velocity correlations 
$\langle\bv(\br,t)\cdot\bv(\br^\prime, t^\prime)\rangle$ in each of the regimes of length and time scales depicted in Fig.~\ref{ltsc}.

%\vspace{.2in}

%\noindent1) achiral regime: $|\br-\br^\prime|\ll L_c$  and $|t-t^\prime|\ll\tau_c$.\vspace{.2in} 
{\it Achiral regime}--- In this regime, the velocity correlations are simply those of an achiral Malthusian flock,  which exhibit long-ranged order \cite{Toner_prl12}, and
whose scaling exponents have been estimated numerically in Ref.~\cite{Chate_prl24}

{\it Linear KPZ regime---}Here, the effects of the $\lambda_{_K}$ non-linearity in the KPZ equation are negligible. Dropping that  term  leaves an analytic solvable linear theory, from which we predict 
that the velocity correlations oscillate rapidly in time and decay algebraically in space and time: 
\bew
\beqn
\langle\bv(\br,t)\cdot\bv(\br',t')\rangle
 \approx v_0^2 
\cos\left[b(t-t')\right]\times\left\{
\begin{array}{ll}
\sqrt{\ee^{-\alpha\gamma}}\left(|\br-\br'|\over 2L_c\right)^{-\alpha}\sep
&\left( \frac{|\br-\br'|}{L_c}\right)^2\gg\frac{|t-t'|}{\tau_c}\sep \frac{|\br-\br'|}{L_c}\gg1\,,\\
\left(|t-t'|\over \tau_c\right)^{-{\alpha\over 2}}\sep
&\frac{|t-t'|}{\tau_c}\gg\left( \frac{|\br-\br'|}{L_c}\right)^2 \sep \frac{|t-t'|}{\tau_c}\gg 1\,,
\end{array}
\right.
\,\label{v_corre2I}
\eeqn
\ew
where the exponent $\alpha$ is non-universal and given by
\beqn
\alpha={D_\phi\over 2\pi\nu}=C_\alpha b^{\eta_{\nu}} \sep \eta_\nu\approx{3\over 5} \ ,
\label{aI}
\eeqn
where $C_\alpha$ is a non-universal constant and $\gamma\approx.57721566...$ is the Euler-Mascheroni constant.

Note that for small chirality $b$, the exponent $\alpha$ will be small, since $\nu(b)$ gets very large in that limit. Hence, the algebraic decay of the equal-time velocity correlations obtained by setting $t=t'$ in \rf{v_corre2I} will be extremely slow; this is what we meant by our earlier cryptic comment that ``some regimes are effectively ordered". Strictly speaking, the order in this regime is ``quasi-long-ranged".

The smallness of the exponent $\alpha$ means the decay of correlations will be so slow that it could easily be mistaken for long-ranged order. We believe that this is what is actually happening in the simulations of references \cite{Liebchen_prl2017} and \cite{Liebchen_epl2022}.

The behavior of the purely temporal Fourier transform of the velocity-velocity correlation function, that is, 
\beq
I(\bR, \omega)\equiv\int_{-\infty}^\infty  \langle\bv(\br,t)\cdot\bv(\br+\bR,t+T)\rangle \, \ee^{\ii\omega T} \,\dd T
\label{idef}
\eeq
is also interesting  in this regime (and experimentally measurable). 
Unlike the equal-time case, the  non-equal-time velocity correlations carry  frequencies $\omega=\pm b$ (see the cosine factor in (\ref{v_corre2I})). This is reminiscent of the spatially periodic density modulation  in crystals. Just as that spatial modulation leads to Bragg peaks in the X-ray scattering, the {\it temporal} periodicity of our system leads to ``Bragg peaks" in the temporal Fourier transform of the velocity correlation at frequencies
%k?
$\omega=\pm b$:
\beqn
I(\bR,\omega)\propto
|\dw|^{\alpha/2-1}
\sep |\dw|\ll{\nu/R^2} \,,
\label{I(w)}
\eeqn
where $\dw\equiv\omega\pm b$ is the difference between the frequency $\omega$ and its value $\pm b$ at the nearest peak. That is, take the ``$+$" sign for $\omega$ near $-b$, and the ``$-$" sign for $\omega$ near $+b$.

This divergence as $\dw\to0$ is cut off very close to the Bragg peaks (specifically, once $|\dw|\lesssim\tau_{_{NL}}^{-1}$) by the more rapid decay of correlations in the non-linear regime, rounding the peak.

%\vspace{.2in}

{\it Nonlinear KPZ regime---}
%: $L_{NL}\ll |\br-\br'| \ll L_v$ and $|t-t'| \ll\tau_{v}$, or $|\br-\br'| \ll L_v$ and $\tau_{NL}\ll |t-t'| \ll\tau_{v}$.
%\vspace{.2in}In this regime, 
The effects of the $\lambda_{_K}$ non-linearity in the KPZ equation now dominate. The mean squared difference between the phases at widely separated spatial and temporal positions $(\br, t)$ and $(\br^\prime, t^\prime)$ are given by (\ref{phi_correlI_phi}).

 If the fluctuations of the field $\phi$ were Gaussian, then  (\ref{phi_correlI_phi})  would imply
 %into this equality we get
% \bew
% \beqn
% \langle\bv(\br,t)\cdot\bv(\br',t')\rangle=v_0^2\cos\left[b(t-t')\right]\exp\left[-\left(A_\phi\over 2\right) |\br-\br'|^{2\chi}\cG_\phi\left(|t-t^\prime|\over|\br-\br'|^z\right)\right]\,.
 %\label{gaussianI1}
 %\eeqn
 %\ew
 %The parameter $A_\phi$ can be calculated by arguing that the phase correlations (\ref{phi_correlI_v0}) and (\ref{phi_corre2I}) must connect at the crossovers, for example, at $|\br-\br'|=L_{NL}$, $|t-t'|=0$. This
 %argument gives
 %\beqn
 %\label{eq:Aphi}
 %A_\phi=2\alpha L_{_{NL}}^{2\chi}\ln\left(L_{NL}\over L_c\right)
 %\propto b^{-{16\over 5}}\exp\left[\left(b^{-3.8}\right)\times{\rm constant}\right]\,.
% \nn\\
% \eeqn
% Again focusing on 
that the equal-time correlation of the velocity would be
%, we'd now have:
 %\beqn
 %\langle\bv(\br,t)\cdot\bv(\br',t)\rangle=v_0^2\exp\left[-\left(A_\phi\over 2\right) |\br-\br'|^{2\chi}\right]\,,
 %\label{gaussianI2}
 %\eeqn
%\beqn
%\big\langle\exp\big\{\ii\left[\phi(\br,t)-\phi(\br^\prime,t)\right]\big\}\big\rangle%\nonumber\\
%=\exp\bigg\{-{1\over2}\bigg\langle\left[\bigg(\phi(\br,t)-\phi(\br^\prime%\br?
%, t)\bigg)^2\right]\bigg\rangle\bigg\}=\exp\left(-\left({A_\phi\over2}\right)|\br-%\br'
%\br^\prime|^{2\chi}\right)\,,
%\label{gaussianI}
%\eeqn
%which would imply (
stretched exponentially decaying velocity correlations.
However, since the $\phi$ fluctuations are non-Gaussian, due to the relevant nonlinearity $\lambda_{_K}$ in the KPZ equation, we can say very little, beyond noting that we expect velocity correlations to decay rapidly with increasing  separation $|\br-\br^\prime|$ once that separation is larger than $L_{_{NL}}$.  
%that %\sout{$\langle[\phi(\br,t)-\phi(\br',t)]^2\rangle\gtrsim O(1)$} $A_\phi |\br-\br'|^{2\chi}\gtrsim O(1)$. \sout{Using \rf{phi_correlI}, w}

{\it Vortices unbound regime---}Here, vortices in the phase field $\phi$ disorder the system even more rapidly, causing the 
%: $L_{v}\ll |\br-\br^\prime|$ or $\tau_{v}\ll |t-t^\prime|$.\vspace{.2in}As we discussed earlier,the fact that this regime even exists is a consequence of the fact that, strictly speaking,  the Malthusian chiral flock maps onto the {\it compact} $(2+1)$-dimensional KPZ equation, which allows vortices in the system. We believe, based on the previously discussed analogy between chiral Malthusian flocks and the equilibrium XY model, that the typical spacing between vortices is $L_v$, and that for length scales longer than $L_v$ it is therefore  impossible \cite{KT} to even {\it define} a single-valued phase field $\phi(\br,t)$ over all space $\br$. We also expect the 
velocity correlations to be short-ranged (indeed, to essentially vanish) in this range of length and time scales.

In summary, we have shown that chiral Malthusian flocks in two dimensions belong generically to the KPZ universality. As a result, chirality always disorders a 2D chiral Malthusian flock, leading to short-ranged velocity correlations. We have further shown that for small chirality, the velocity can remain well correlated out to quite long distances. Finally, we have made a number of very detailed predictions for the behavior of these correlation functions that are testable in both simulations and experiments on active chiral systems.

We thank the Max Planck Institute for the Physics of Complex Systems, where the early stage of this work was performed, for their support. L.C. acknowledges support by the National Science Foundation
of China (under Grants No.12274452).

We also thank Tim Halpin-Healy and Mehran Kardar for useful discussions about the current state of our understanding of the KPZ equation, and for calling references \cite{kpzexp4,kpzexp41}  and \cite{kpzexp5} to our attention.  Last but not least, we are grateful to Ananyo Maitra for valuable discussions about prior work on chirality in Malthusian flocks, and for suggesting the name ``time cholesteric".

{\it Note added} - After this work was completed, we learned that the conclusion that chiral Malthusian flocks are described by the KPZ equation had been independently reached by Ananyo Maitra\cite{scoop}, whom we thank  for calling this to our attention.

\end{document}